\documentstyle[12pt,a41,axodraw,epsfig,pstricks,pst-node,cite,amsmath,float]{article}

\newcommand{\gsim}{\raisebox{-0.07cm}{$\,\stackrel{>}{{\scriptstyle
 \sim}}\, $} }

\newcommand\eps{\varepsilon}
\newcommand\ep{\varepsilon}

\newcommand\GeV{\,\mbox{GeV}}

\newcommand{\bq}{\begin{equation}}
\newcommand{\eq}{\end{equation}}
\newcommand\beq{\begin{equation}}
\newcommand\eeq{\end{equation}}
\newcommand\bea{\begin{eqnarray}}
\newcommand\eea{\end{eqnarray}}

\include{epsfig}
\begin{document}
\setlength{\baselineskip}{0.515cm}
\sloppy
\thispagestyle{empty}
\begin{flushleft}
DESY 06--120 \hfill
{\tt hep-ph/0702265}\\
SFB-CPP/07-06\\
February 2007\\
\end{flushleft}

\mbox{}
\vspace*{\fill}
\begin{center}

{\Large\bf Calculation of Massive 2--Loop Operator } 

\vspace{2mm}
{\Large\bf 
Matrix Elements
with Outer Gluon Lines} 

\vspace{4cm}
\large
I. Bierenbaum, J. Bl\"umlein, and S. Klein 

\vspace{1.5cm}
\normalsize
{\it  Deutsches Elektronen--Synchrotron, DESY,}\\
{\it  Platanenallee 6, D-15738 Zeuthen, Germany}
\\

\end{center}
\normalsize
\vspace{\fill}
\begin{abstract}
\noindent
Massive on--shell operator matrix elements and self-energy diagrams with 
outer gluon lines are calculated analytically at $O(\alpha_s^2)$, using 
Mellin--Barnes integrals and representations through generalized 
hypergeometric functions. This method allows for a direct evaluation 
without decomposing the integrals using the integration-by-parts method.  
\end{abstract}

\vspace{1mm}
\noindent

\vspace*{\fill}
\noindent
\newpage
\section{Introduction}
%

\vspace{1mm}\noindent
In the asymptotic region $Q^2 \gg m^2$, the heavy flavor contributions
to the deeply inelastic structure functions can be obtained from the 
corresponding massive operator matrix elements and the light flavor
Wilson coefficients \cite{Buza:1995ie}. The massless Wilson coefficients 
for deeply inelastic scattering are known up to 3--loop order 
\cite{WILS2,WILS3}. The heavy flavor contributions were calculated to
next-to-leading order in \cite{HEAV} semi-analytically. A fast numerical 
implementation was given in \cite{Alekhin:2003ev}. Complete analytic 
results were derived only for the limit $Q^2 \gg m^2$ for the structure 
function $F_2^{Q\overline{Q}}(x,Q^2)$ to $O(\alpha_s^2)$ \cite{Buza:1995ie}
and $F_L^{Q\overline{Q}}(x,Q^2)$ to $O(\alpha_s^3)$ 
\cite{Blumlein:2006mh}.
In both cases, the $O(\alpha_s^2)$ massive operator matrix elements are
required. The asymptotic contributions cover all logarithmic and the constant 
terms, while contributions of $O((m^2/Q^2)^k),~k \geq 1$ are not 
contained. In the case of the structure function $F_2(x,Q^2)$, these
terms yield a very good description already in the region $Q^2 \gsim 20 
\GeV^2$, while for $F_L(x,Q^2)$ this approximation only holds at large 
scales $Q^2 \gsim 1000 \GeV^2$. Since the heavy flavor contributions to 
the structure functions amount to 20--40~\% in the small $x$ region, 
cf.~\cite{STRAT}, and the scaling violations of these terms differ
from that of the light parton contributions, their knowledge is essential 
for precision measurements of the QCD scale $\Lambda_{\rm QCD}$ in singlet
analyses.

In this letter we address a new compact calculation of the genuine 2--loop 
scalar integrals contributing to the massive operator matrix elements with
outer gluon lines, based on the Mellin-Barnes technique \cite{MB1,MB2,MB3} 
and using 
representations through generalized hypergeometric functions \cite{HGF}. This approach
allows to thoroughly avoid the use of the integration-by-parts method \cite{IBP}, which 
keeps the contributing number of terms low and yields very compact results.
Moreover, we work in Mellin space to use the appropriate symmetry of the
problem leading to further compactification. The complete calculation of the
asymptotic heavy flavor Wilson coefficients will be presented elsewhere
\cite{BBK3}. In the following, we will outline  the principal method and
present then the results for the seven contributing two--loop integrals in 
terms of nested harmonic sums \cite{HSUM1,HSUM2}. Some of the special sums 
needed are listed in the appendix.   

\section{The Method}
%
\vspace{1mm}\noindent
The massive 2--loop diagrams considered are shown in Figure~1. 
\begin{figure}[h]
\begin{center}
   \epsfig{file=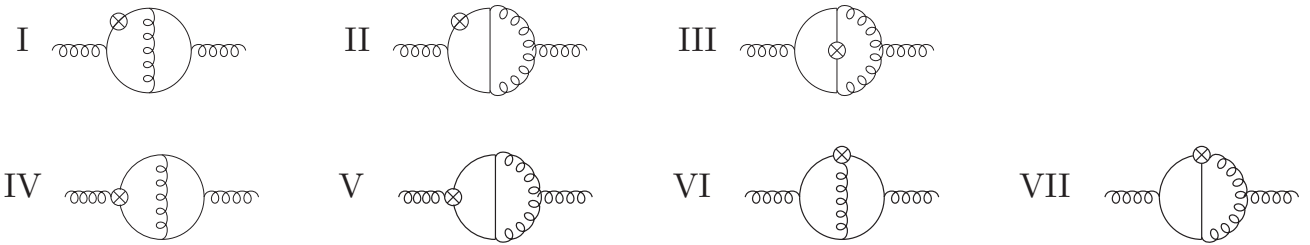,width=\linewidth}
\end{center}
\caption{\small \sf
Genuine 2-loop diagrams contributing to the massive operator matrix 
elements.}
\label{fig:diag}
\end{figure}
The diagrams contain either three or four massive 
lines. The $\otimes$--symbol in Figure~1 denotes the operator insertion of the
corresponding local quark-gluon operators, see Figure~2.
\begin{figure}[h]
\begin{center}
    \epsfig{file=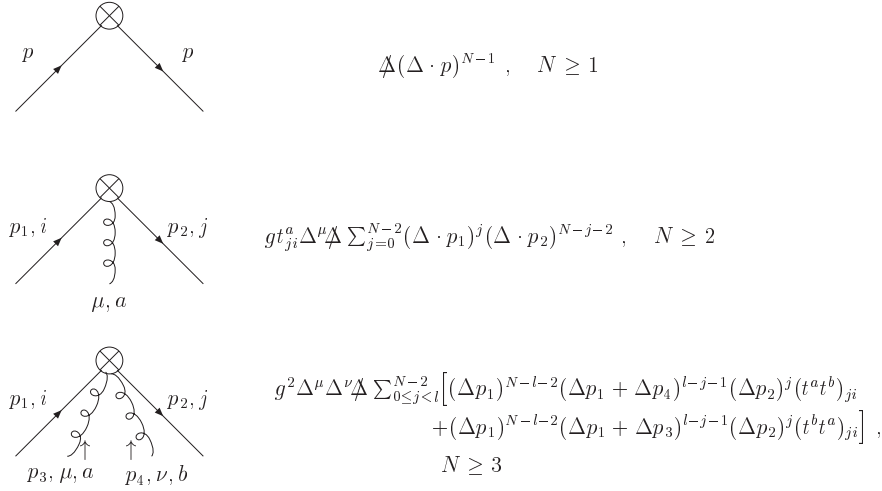,height=6.5cm}
\end{center}
\caption{\small \sf
QCD Feynman rules for the composite local operator insertions.}
\label{fig:op-feuynr}
\end{figure}

The diagrams  can be decomposed into a *--product as
described in Figure~3.
We follow  the calculation of Ref.~\cite{MB3}, now generalized from
massless self--energy diagrams to massive 
operator matrix elements. 
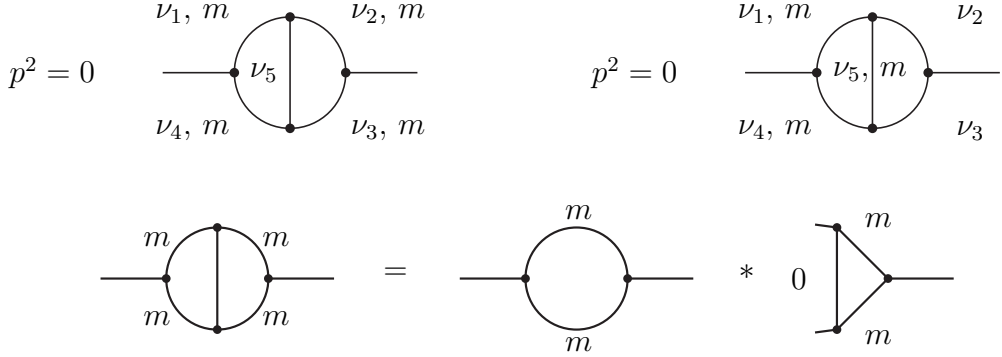
\begin{figure}[ht]
\begin{center}
\SetScale{0.6} 
\SetWidth{1}
\hspace{2cm}
\begin{picture}(30,30)(0,0)
  \Line(-80,0)(-35,0)
  \Line(35,0)(80,0)
  \BCirc(0,0){35}
  \Line(0,35)(0,-35)
  \Vertex(-35,0){3}
  \Vertex(35,0){3}
  \Vertex(0,35){3}
  \Vertex(0,-35){3}
  \Text(-37,-22)[]{\(\nu_4,\, m\)}
  \Text(37,-22)[]{\(\nu_3,\, m\)}
  \Text(-37,22)[]{\(\nu_1,\, m\)}
  \Text(37,22)[]{\(\nu_2,\, m\)}
  \Text(-10,0)[]{\(\nu_5\)}
  \Text(-90,0)[]{\(p^2=0\)}
\end{picture}
\hspace{5cm}
\SetScale{0.6} 
\SetWidth{1}
\hspace{1cm}
\begin{picture}(30,30)(0,0)
  \Line(-80,0)(-35,0)
  \Line(35,0)(80,0)
  \BCirc(0,0){35}
  \Line(0,35)(0,-35)
  \Vertex(-35,0){3}
  \Vertex(35,0){3}
  \Vertex(0,35){3}
  \Vertex(0,-35){3}
  \Text(-37,-22)[]{\(\nu_4,\, m\)}
  \Text(37,-22)[]{\(\nu_3\)}
  \Text(-37,22)[]{\(\nu_1,\, m\)}
  \Text(37,22)[]{\(\nu_2\)}
  \Text(-1,0)[]{\(\nu_5,\,  m\)}
  \Text(-90,0)[]{\(p^2=0\)}
\end{picture}
\\[4em]
  \begin{tabular}{cccc}
    \SetScale{0.55}
    \SetWidth{1.5}
    \hspace{-1.2cm}
    \begin{picture}(50,30)(0,0)
      \Line(70,0)(115,0)
      \Line(185,0)(230,0)
      \BCirc(150,0){35}
      \Line(150,35)(150,-35)
      \Vertex(115,0){3}
      \Vertex(185,0){3}
      \Vertex(150,35){3}
      \Vertex(150,-35){3}
      \Text(60,15)[]{\(m\)}
      \Text(105,15)[]{\(m\)}
      \Text(60,-15)[]{\(m\)}
      \Text(105,-15)[]{\(m\)}
    \end{picture}
    &
    \hspace{2.8cm}
    =
    &
    \SetScale{0.55}
    \SetWidth{1.5}
    \begin{picture}(70,30)(0,0)
      \Line(0,0)(45,0)
      \Line(115,0)(160,0)
      \BCirc(80,0){35}
      \Vertex(45,0){3}\Vertex(115,0){3}
      \Text(45,25)[]{\(m\)}
      \Text(45,-25)[]{\(m\)}
    \end{picture}
    &\qquad
    $\ast$
    \SetScale{0.55}
    \SetWidth{1.5}
    \begin{picture}(90,30)(0,0)
      \Line(70,0)(115,0)
      \Line(35,35)(35,-35)
      \Line(70,0)(35,35)
      \Line(70,0)(35,-35)
      \Line(35,35)(20,37)
      \Line(35,-35)(20,-37)
      \Vertex(70,0){3}
      \Vertex(35,35){3}
      \Vertex(35,-35){3}
      \Text(5,0)[]{\(0\)}
      \Text(35,22)[]{\(m\)}
      \Text(35,-22)[]{\(m\)}
    \end{picture}
    \\&&&\\
  \end{tabular}
\end{center}
\caption{\small \sf The massive two--loop two--point diagrams with three 
and 
four fermion masses. The graphs are generated by inserting the three--point 
function, written as a double Mellin--Barnes integral, into the massive 
two--point function, i.e. performing the $\ast$-operation defined in the 
context of the Lie--algebra of Feynman diagrams \cite{CK}. The effect of 
this 
insertion is only given by a modification of the exponents of the 
two--point function, cf.~\cite{MB3}.}
\label{fig:master_2l_2p}
\end{figure}

\vspace{1mm}\noindent
Also in this case the above decomposition of diagrams 
can be achieved by applying the Mellin--Barnes representation~\footnote{
In his original contribution, 
Barnes notes that the contour integral representations 
(\ref{eqMB}) and those for more complicated integrands date back to 
Pincherle
\cite{PINCH} Mellin \cite{MB2} and Riemann \cite{RIE}, cf also \cite{KF}.}
\begin{eqnarray}
\label{eqMB}
\frac{1}{(A_1+A_2)^{\nu}}
&=&
\frac{1}{2\pi i} \int_{\gamma-i\infty}^{\gamma+i\infty}
      d\sigma \:
      A_1^{\sigma}A_2^{-\nu-\sigma}
      \frac{\Gamma(-\sigma)\Gamma(\nu+\sigma)}{\Gamma(\nu)}~.
\end{eqnarray}

Let us consider Diagram~I as an example. The corresponding gluing-product 
is depicted in Figure~4.
Applying the Feynman-parameterization to the 2-point function yields
\begin{eqnarray}
{I}^{(1,2)} &=& 
\frac{\Gamma(\nu_{14})}{\Gamma(\nu_1) 
\Gamma(\nu_4)}~{(m^2)^{\nu_{14}-D/2}}{(4\pi)^2}
\int_0^1 dx_1 dx_2 x_1^{\nu_1-1} x_2^{\nu_4-1} 
\delta(x_1+x_2-1) \nonumber
\end{eqnarray}\begin{eqnarray} 
& & \times
\int \frac{d^D k_1}{(2\pi)^{D}} 
\frac{(\Delta.k_1)^{N-1}}{(x_1 k_1^2 +x_1 m^2 + x_2(k_1-p)^2 +x_2 
m^2)^{\nu_{14}}}~.
\end{eqnarray}

\vspace{1mm}\noindent
Here, $\Delta$ denotes a light-like vector with $\Delta^2 = 0$, and 
$D =4 - 2\eps$. $\nu_i$ is 
the integer power of the respective propagator and $\nu_{ij \ldots} = 
\nu_i + \nu_j + \ldots$~.
The calculation is performed in the $\overline{\rm MS}$ scheme and we 
factor out $S_\eps = \exp[\eps (\ln(4\pi) - \gamma_E)]$ for each 
loop.~\footnote{All integrals are normalized to contain 
no mass-scale or factors of $2\pi$ in the final result.} 
One shifts $k_1 \rightarrow k_1 + x_2 p$ and the numerator term is 
decomposed as
\begin{eqnarray}
(\Delta.k_1 + x_2 \Delta.p)^N = \sum_{l=0}^N \binom{N}{l} (\Delta.k_1)^l 
(x_2 \Delta.p)^{N-l}~.
\end{eqnarray}
All integrals over $k_1$ with $(\Delta.k_1)^l,~l \geq 1$ vanish.
\begin{figure}[h]
\begin{center}
    \epsfig{file=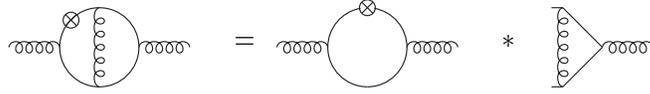,width=0.5\linewidth}
\end{center}
\caption{\small \sf
Diagram insertion for graph~I.}
\label{fig:diag1}
\end{figure}

\vspace{2mm}\noindent
One obtains
\begin{eqnarray}
{I}^{(1,2)} &=& (\Delta.p)^{N-1} (4\pi)^\eps
\frac{\Gamma(\nu_{14}-D/2)\Gamma(\nu_4+N-1)}{ 
\Gamma(\nu_4)\Gamma(\nu_{14}+N-1)}~.
\end{eqnarray}

In the next step, the three-point function is inserted in a similar way 
as in \cite{MB3}, 
since a part of the propagators are massive but no  
operator insertion occurs.
The result for diagram~I for $\nu_i =1, \forall i$, can be expressed by 
a double Mellin--Barnes 
integral as
\begin{eqnarray}
\label{G1_11111_st}
  {I}_{1} 
  &=& \frac{(4\pi)^{2\eps}}{(2\pi 
)^2}\frac{(\Delta.p)^{N-1}}{\Gamma(1-2\eps)}
  \nonumber\\
  & & \times\int_{\gamma_1-i\infty}^{\gamma_1+i\infty} d\sigma
  \int_{\gamma_2-i\infty}^{\gamma_2+i\infty} d\tau \:
  \frac{\Gamma(-\sigma)\Gamma(\sigma+1)\Gamma(-\sigma+N)}{\Gamma(-\sigma+1)}
  \Gamma(-\tau)\Gamma(\tau+1)\nonumber
  \nonumber\\
  & & \times
  \frac{\Gamma(\sigma+\tau+1)
\Gamma(\sigma+\tau+1+\eps)\Gamma(-\sigma-\tau-2\eps)
\Gamma(-\sigma-\tau +\eps)}
      {\Gamma(\sigma+\tau+2)\Gamma(-\sigma-\tau+1+N)}~.
\end{eqnarray}
The other six 2-loop integrals obey a similar 
representation. For fixed values of $N$, one may calculate the 
Mellin--Barnes integrals using the {\tt mathematica}--package {\tt MB} by
M.~Czakon \cite{Czakon:2005rk}, which yields numerical values. They were 
given in Ref.~\cite{BBK2}, Table~1.~\footnote{Note that in \cite{BBK2}
the spherical factor $S_\ep^2$ has not been factored out.} 

For the solution of one of the  Mellin-Barnes integrals, relations like
\begin{eqnarray}
\label{generalized_Barnes}
\lefteqn{
\hspace{-14cm}
\frac{1}
     {2\pi i} 
     \int_{\gamma-i\infty}^{\gamma+i\infty} ds
\frac{\Gamma(a+s)\Gamma(b+s)\Gamma(d-a-b-s)\Gamma(e-c+s)\Gamma(-s)}
     {\Gamma(e+s)}
} \\
\hspace{4cm} 
=
\frac{\Gamma(e-c)\Gamma(a)\Gamma(b)\Gamma(d-a)\Gamma(d-b)}
     {\Gamma(e)\Gamma(d)}
     ~_3F_2[a,b,c;d,e;1]~, \nonumber
\end{eqnarray}
cf.~\cite{HGF}, are used. The second integral is performed by applying the 
residue theorem. One obtains 
\begin{eqnarray}
\label{eqA}
{I}_{1} &=&
(\Delta \cdot p)^{N-1}\: (4\pi)^{2\ep}
\frac{\Gamma(N+1)}{\Gamma(1-2\eps)}
\:\sum_{k=0}^{\infty} \sum_{j=0}^{\infty} 
\frac{\Gamma(k+1)}{\Gamma(k+2+N)}\nonumber
\\
& &\times
\Biggl[
\Gamma(\eps)\Gamma(1-\eps)
\frac{\Gamma(j+1-2\eps)\Gamma(j+1+\eps)}{\Gamma(j+1-\eps)\Gamma(j+2+N)} 
\frac{\Gamma(k+j+1+N)}{\Gamma(k+j+2)} 
\nonumber\\
& &
+
\Gamma(-\eps)\Gamma(1+\eps)
\frac{\Gamma(j+1+2\eps)\Gamma(j+1-\eps)}{\Gamma(j+1)\Gamma(j+2+\eps+N)} 
\frac{\Gamma(k+j+1+\eps+N)}{\Gamma(k+j+2+\eps)} 
\Biggr]~.
\end{eqnarray}
One first performs the $\eps$--expansion to the desired order.
The infinite sums in (\ref{eqA}) can be expressed through Mellin-type integral 
representations, partly using differential operators in the remaining 
summation and outer parameters. In some cases the starting values 
$\{k,j\}=0$ 
need a separate treatment. The corresponding Mellin-integrals finally 
result into weighted harmonic sums. The calculations were
coded in {\tt MAPLE}. 
For fixed values of 
$N$, simpler procedures are 
obtained which lead to analytic expressions for the expansion coefficients
in $\eps$, cf. Table~2, Ref.~\cite{BBK2}. 
\footnote{In case of diagram ${\rm I}_a$ and II checks 
could also be performed by {\tt nestedsums} \cite{NESU} at fixed $N$.} 

Except for diagrams VI, VII one 
may 
calculate the integrals in the above way for general values of $N$.
   
Explicit representations for all diagrams could be derived using 
generalized hypergeometric functions for all diagrams. 
As an example, let us consider diagram VI,
 \begin{eqnarray}
  I_6&=&\int\!\!\frac{dk_1}{(2 \pi)^{D}}\int\!\!\frac{dk_2}{(2 \pi)^{D}}
       \frac{
(\Delta k_1)^{N-1}\!-\!(\Delta k_2)^{N-1}}
       {(\Delta k_1\!-\!\Delta k_2)}
       \nonumber\\ & & \times (4\pi)^4~
       \frac{(m^2)^{1+2\eps}}{(k_1^2\!-\!m^2)((k_1\!-\!p)^2\!
       -\!m^2)(k_2^2\!-\!m^2)((k_2\!-\!p)^2\!
       -\!m^2)^2(k_2\!-\!k_1)^2}
       \nonumber \\
     &=&(\Delta p)^{N-2}\Gamma(1+2\eps) (4\pi)^{2 \ep}
        \frac{2\pi}{N\sin(-\pi\eps)}
        \sum_{j=1}^{N}\Biggl\{\binom{N}{j}(-1)^j+\delta_{j,N}\Biggr\}
        \label{I6erg} \\
     && \times\Biggl\{
        \frac{\Gamma(j)\Gamma(j+1+\eps)}
        {\Gamma(j+2+2\eps)\Gamma(j+1-\eps)}
       -\frac{B(1+\eps,1+j)}{j}~ 
        _3F_2\left[1+2\eps,-\eps,j+1; 1,j+2+\eps; 1
	\right]\Biggr\}~. \nonumber
 \end{eqnarray}
After performing the expansion in $\eps$, the remaining sums can be 
carried 
out by suitable integral representations.
Some of the sums required are listed in the appendix.
\section{Results}
%

\vspace{1mm}\noindent
All diagrams obey representations in terms of weighted harmonic sums in 
the present case.
It even turns out that only single harmonic sums contribute. 
Not all the diagrams I--VII are independent. Due to the
massless outer lines of the diagrams,  graphs IV and 
V can be expressed  
in terms of diagrams ${\rm I}_a$ and II. Here ${\rm I}_a$ corresponds to 
the case where all scalar propagators have power one, i.e.
\begin{eqnarray}
{\rm IV}  &=& \left [ 1+(-1)^N\right] \frac{1}{\Delta . p}\times {\rm I}_a 
\\
{\rm V}   &=& \left [ 1+(-1)^N\right] \frac{1}{\Delta . p} \times {\rm II} 
\end{eqnarray}
For diagram~III, even a closed expression for general values of $\eps$ 
can be derived,
 \begin{eqnarray}
I_3&=&\int\!\!\!\frac{dk_1}{(2\pi)^{D}}\int\!\!\!\frac{dk_2}{(2\pi)^D}
       \frac{(4\pi)^4 (\Delta k_1\!-\!\Delta k_2)^{N-1}\quad 
(m^2)^{1+2\eps}}
       {(k_1^2\!-\!m^2)((k_1\!-\!p)^2\!-\!m^2)((k_1\!-\!k_2)^2\!-\!m^2)k_2^2(k_2\!-\!p)^2}
       \nonumber\\
     &=&(\Delta p)^{N-1}\Gamma(1+2\eps) (4 \pi)^{2\ep}
        \frac{\pi}{\sin(-\pi\eps)}
        \frac{1-(-1)^N}{N(N+1)}\frac{\Gamma(N+1+\eps)
        \Gamma(N+1)}{\Gamma(N+2+2\eps)\Gamma(N+1-\eps)}
        \label{I3erg}\\
     &=&(\Delta p)^{N-1}\Gamma(1+2\eps) (4 \pi)^{2\ep}
        \frac{1-(-1)^N}{N(N+1)^2}\Biggl[-\frac{1}{\eps}
        +\frac{2}{N+1}\Biggr]+O(\eps) \nonumber.
 \end{eqnarray}
We summarize the results for the independent diagrams I-III,VI,VII in 
Table~1, expanding to $O(\eps^0)$ in the $\overline{\rm MS}$--scheme.
The expressions in terms of harmonic sums were derived exploiting their
algebraic relations~\cite{ALGEBRA}.

\begin{table*}[htb]
\caption{The analytic results for graphs I to VII for general values of N, with all $\nu_i = 1$, Ib: $\nu_1 = 2$.}
\label{table:results3}
\newcommand{\m}{\hphantom{$-$}}
\newcommand{\cc}[1]{\multicolumn{1}{c}{#1}}
\renewcommand{\arraystretch}{2.3} 
\begin{tabular}{ll}
\hline\hline
$\rm{Ia}$            & \m $\displaystyle 
\frac{{S}_1^2({N})+3{S}_2({N})}{2{N}({N}+1)}$ \\
$\rm{Ib}$            & \m $\displaystyle 
\frac{{S}_1({N})-3 
S_2(N)/2-S_{1}^2(N)/2}{{N}({N}+1)(N+2)}- 
\frac{1}{(N+1)^2(N+2)}$ \\
$\rm{II}$           & \m $\displaystyle 
\frac{{S}_1({N})}{{N}({N}+1)}\left(-\frac{1}{\eps} \right)
+2\frac{{S}_1({N})}{{N}({N}+1)^2}+\frac{{S}_1^2({N})-{S}_2({N})}{2{N}({N}+1)}$  
\\
$\rm{III}$          & \m $\displaystyle 
\frac{[1-(-1)^{{N}}]}{{N}({N}+1)^2}\left(-\frac{1}{\eps}+\frac{2}{({N}+1)}
\right)$  \\
$\rm{VI}$            & \m $\displaystyle \frac{4}{N} \left[S_2(N) - \frac{S_1(N)}{N} \right]$ \\ 
$\rm{VII}$            &  
\m $\displaystyle 
\left[\frac{(-1)^N -1}{N^2(N+1)} + \frac{2S_1(N)}{N(N+1)}\right]\left(-\frac{1}{\eps} \right)$
\\ & \m $\displaystyle
+\left[2\frac{(-1)^N -1}{N^2(N+1)^2} + 
\frac{S_1^2(N)-S_2(N)+2S_{-2}(N)}{N(N+1)}+\frac{2(3N+1)S_1(N)}{N^2(N+1)^2}\right]$\\
[3mm]
\hline 
\hline
\end{tabular}\\[2pt]
\end{table*}

\noindent
In terms of (single) harmonic sums 
\begin{equation}
S_a(N) = \sum_{k=1}^N \frac{({\rm sign}(a))^k}{k^{|a|}}~,
\end{equation}
the final expressions for the genuine 2--loop 
diagrams turn out to be extraordinarily simple. 
Due to the fact that we were not applying the integration-by-parts method, 
also the intermediary results remained rather compact. 
The results may be translated into $x$-space by inverse Mellin 
transformation
using the Tables in \cite{HSUM1}. On the other hand, one may work within 
the Mellin space representation continuing the expressions to complex values 
of $N$ as described in \cite{ANCONT} and construct the respective 
observables in analytic form. The inverse Mellin transformation is then 
performed by a single numeric integral.

The diagrams 
are applied to express the unpolarized massive 2--loop operator matrix 
elements \cite{Buza:1995ie,BBK3}. As shown in 
\cite{Blumlein:2006mh,BBK3},
nested harmonic sums will contribute in the physical result 
but they are either due to one-loop insertions into one-loop 
diagrams or the four-propagator contributions to some of the topologies 
discussed here, which are obtained due to cancellation of numerator and
denominator terms.

\section{Appendix}
%
\vspace{1mm}\noindent
We list some special sums which are typical for classes of sums to be 
derived for the present calculation.
\begin{eqnarray}
\sum_{k=1}^\infty
\sum_{i=1}^\infty \frac{S_1(k+i+N)}{(k+i) (N+k) k} 
&=& \frac{1}{2} \sigma_1^2 \frac{S_1(N)}{N} + 2 \frac{S_{1,1}(N)}{N^2}
- \frac{\zeta_2}{2} \frac{S_1(N)}{N} - \frac{2 \zeta_3}{N} \\
\sum_{i=1}^\infty \frac{B(N,i)}{i} &=& 
\zeta_2 - S_2(N-1)\\
\sum_{i=1}^\infty \frac{B(N+1,i)}{(N+i)} &=& (-1)^N \left[2 S_{-2}(N) + 
\zeta_2 \right]\\
\sum_{i=1}^\infty \frac{B(N,i)}{(N+i+1)^2} &=&
\frac{(-1)^N}{N(N+1)}\left[ 2 S_{-2}(N) + \zeta_2\right] + 
\frac{N-1}{N(N+1)^3}\\
\sum_{i=1}^\infty \frac{B(N+1,i) S_1(i)}{(N+i)} &=& 
\frac{\zeta_2-S_2(N)}{N} 
+ (-1)^N \Biggl[\zeta_3 + S_{-3}(N) 
- 2 \frac{S_{-2}(N)}{N} 
\nonumber \\ & &
+ 2 S_{1,-2}(N) - \frac{\zeta_2}{N} + \zeta_2 
S_1(N)\Biggr]\\
\sum_{k=0}^{N-1}\binom{N-1}{k} (-1)^k \frac{S_1^2(k+2)}{k+2}
&=&
- \frac{2N+1}{N^2 (N+1)^2} S_1(N) + \frac{N^3+6N^2+6N+2}{N^3(N+1)^3}
\\
\sum_{k=0}^L \binom{L+1}{k} \frac{(-1)^k}{(N-L+k)^2}
&=& B(N-L,L+2) \left[ S_1(N+L) - S_1(N-L-1)\right]~.
\end{eqnarray}
Here the symbol $\sigma_1$ and Euler's Beta-function $B(a,b)$ are defined 
by
\begin{eqnarray}
\sigma_1 &=& \lim_{N \rightarrow \infty} S_1(N) \\
B(a,b) &=& \frac{\Gamma(a) \Gamma(b)}{\Gamma(a+b)}~.
\end{eqnarray}


\end{document}